# Complementary logic operation based on electric-field controlled spin-orbit torques

Seung-heon Chris Baek[1,2,*], Kyung-Woong Park[1,3,*], Deok-Sin Kil[3], Kyung-Jin Lee[4,5] and Byong-Guk Park[1,**]

[1]*Department of Materials Science and Engineering and KI for Nanocentury, KAIST, Daejeon 34141, Korea*

[2]*School of Electrical Engineering, KAIST, Daejeon 34141, Korea*

[3]*Research and Development Division, SK Hynix Semiconductor, Inc., Gyeonggi-do 17336, Korea*

[4]*Department of Materials Science and Engineering, Korea University, Seoul 02841, Korea*

[5] *KU-KIST Graduate School of Converging Science and Technology, Korea University, Seoul 02841, Korea*

*These authors equally contributed to this work.

**Corresponding emails: bgpark@kaist.ac.kr (B.-G.P.).

Spintronic devices as alternatives to traditional semiconductor-based electronic devices attract considerable interest as they offer zero quiescent power, built-in memory, scalability, and reconfigurability. To realize spintronic logic gates for practical use, a complementary logic operation is essential but still missing despite a recent progress in spin-based logic devices. Here, we report the development of a complementary spin logic device using electric-field controlled spin-orbit torque (SOT) switching. In heavy metal/ferromagnet/oxide structures, the critical current for SOT-induced switching of perpendicular magnetization is efficiently modulated by an electric field via voltage-controlled magnetic anisotropy (VCMA) effect in a non-volatile manner. Moreover, the polarity of the VCMA is tuned by the modification of oxidation state at the ferromagnet/oxide interface. This allows us to fabricate both "n-type" and "p-type" spin logic devices and to enable a complementary logic operation, paving the way for the development of non-volatile and reconfigurable logic devices.

Spintronic devices for information processing have attracted considerable interest owing to their potential advantages of improved computational capability and reduced power consumption [1-3]. Up to now, spin-based logic devices have been developed in two different approaches. The first is to add the electron spin as a variable parameter into existing semiconductor devices [4], such as the spin field-effect transistor (FET) proposed by Datta and Das [5]. Despite recent successes in the operation of the spin FET with a semiconducting channel [6-8], further developments are required to improve the signal at room temperature and to secure the device scalability. The second approach is to utilize metallic structures in which a logic operation is performed by using domain wall motion [9-12], magnetization switching [13,14], and spin wave propagation [15]. These metal-based spintronic devices are advantageous as they operate at room temperature. However, the screening effect of metals inhibits implementing electric-gating-induced field-effect functions, which are indispensable for complex logic operations. More importantly, a spintronic device performing complementary logic operation analogous to the conventional CMOS has not been developed yet.

In this work, we experimentally demonstrate a spin logic device capable of performing complementary logic operation which is based on the spin-orbit torque (SOT) [16-21] and its controllability using an electric field [22,23]. The SOT is a spin-transfer torque originating from the spin-orbit coupling in heavy metal (HM)/ferromagnet (FM)/oxide structures and enables magnetization switching by an in-plane current. The critical current of the SOT-induced switching can be modulated by a perpendicular gate voltage (see Fig. 1a) which induces the change in the magnetic anisotropy of the trilayer structures via voltage-controlled magnetic anisotropy (VCMA) effect [24-32]. Therefore, the combination of SOT and VCMA allows us to control our device to be either switchable or non-switchable for a given input clock current. This is the first working principle of our spin logic device. The second principle is based on the finding that the polarity of the electric-field effect on the magnetic anisotropy (or VCMA effect) can be reversed by modulating oxidation state at the FM/oxide interface. As we show below, for an under-oxidized interface, a positive (negative) gate voltage decreases (increases) the magnetic anisotropy, whereas for an over-oxidized interface, the reversed VCMA effect occurs. This opposite dependence of VCMA effect on the sign of gate voltage allows us to form an "n-type" or a "p-type" spin logic device by controlling the oxidation state. Having these two types of devices (i.e., n-type and p-type), we realize the complementary functionality of spin logic devices as we demonstrate below.

**Electric-field controlled spin-orbit torque switching**

In order to prove the working principles in our devices, we carry out SOT-switching experiment combined with the application of a gate voltage for Ta(5 nm)/CoFeB(1 nm)/MgO(1.6 nm)/AlO$_x$(1.8 nm) Hall bar structures, where perpendicular magnetic anisotropy (PMA) is established in the CoFeB (see Fig. 1a and Methods). We integrate the gate oxide (ZrO$_2$) and electrode (Ru) on top of a square-shaped CoFeB/MgO/AlO$_x$ island for gating. We vary the oxidation time ($t_{ox}$) for AlO$_x$ formation from 25 sec to 125 sec in order to control the oxidation state at the CoFeB/MgO interface and to modify the dependence of magnetic anisotropy on the voltage polarity. In order to estimate VCMA effect, we measure the anomalous Hall resistance ($R_{xy}$) as a function of the in-plane magnetic field ($B_x$) under a gate voltage ($V_G$), where $R_{xy}$ represents the z-component of magnetization. Analysing the decrease in $R_{xy}$ with increasing $B_x$ allows us to estimate the perpendicular anisotropy field ($H_K$). We find that the $H_K$ maximizes at $t_{ox}$ = 75 sec, referred to as the optimal oxidation time $t_{ox,op}$, and slightly decreases for $t_{ox}$ either longer or shorter than $t_{ox,op}$ (See Supplementary Note 1). Figures 1b – 1d show the dependence of magnetic anisotropy on the gate voltage polarity for samples with different oxidation times. The sample with $t_{ox}$ = 25 sec (< $t_{ox,op}$) shows that the $B_x$–dependent decreasing rate of $R_{xy}$ is larger for $V_G$ of +22 V than for $V_G$ of -20 V (Fig. 1b), indicating that a negative (positive) $V_G$ increases (decreases) $H_K$. Remarkably, the VCMA effect is reversed in its sign for the sample with $t_{ox}$ = 125 sec (> $t_{ox,op}$; Fig. 1d). On the other hand, the VCMA effect is negligible in the sample with $t_{ox}$ = 75 sec (= $t_{ox,op}$; Fig. 1c). The dependence of VCMA effect on $t_{ox}$ is summarized in Fig. 1e. This dependence can be explained by the modification in the oxidation state at the CoFeB/MgO interface and the resultant change in PMA [33]. The application of an electric voltage induces the relocation of oxygen ions either towards or away from the interface depending on its polarity [34], which modulates $H_K$ accordingly. For example, in the under-oxidation condition ($t_{ox}$ < $t_{ox.op}$), a negative bias supplies oxygen ions to the interface that is initially at the oxygen-deficient state, and consequently increases $H_K$. The opposite VCMA effect occurs in the over-oxidation condition ($t_{ox}$ > $t_{ox.op}$) for which a negative bias decreases $H_K$. This voltage-induced change in oxidation state is supported by a non-volatile behaviour of the VCMA effect; $H_K$ remains the same even after $V_G$ is removed (Supplementary Note 2).

We next investigate the effect of $V_G$ on SOT-induced switching performance. We note that in all SOT switching experiments, we apply an external in-plane field along the current direction (+$x$ direction) for the deterministic switching. We note that this external field can be replaced by an internal exchange bias field [35,36]. The critical current ($I_C$) for SOT-induced magnetization switching is known to be proportional to $H_K$ [37] so that $V_G$ is able to control $I_C$ through the modification of $H_K$. As expected, the SOT-induced switching experiment under $V_G$ for the samples with different $t_{ox}$'s demonstrates a clear dependence of $I_C$ on $V_G$ (Figs. 1f – 1h). The absolute value of $I_C$ of the under-oxidized sample ($t_{ox}$ = 25 sec) is 6.5 mA for $V_G$ of +24 V, which is noticeably smaller than 8.8 mA for $V_G$ of -24 V (Fig. 1f). The over-oxidized sample ($t_{ox}$ = 125 sec) shows the opposite behaviour; |$I_C$| for a negative $V_G$ is smaller than that for a positive $V_G$ (Fig. 1h). On the other hand, the sample with $t_{ox}$ = 75 sec shows a negligible

modulation in $I_C$ by $V_G$ (Fig. 1g). Figure 1e shows that the modulation of the $I_C$ is closely correlated with that of the $H_K$. We note, however, that the change in $I_C$ of ~30% is much greater than that in $H_K$ of ~2%, which implies the VCMA may not be the only source causing the electric-field controlled SOT.

These results presented in Fig. 1 demonstrate two working principles required for the realization of the spin logic devices; the first is that $I_C$ for SOT-induced switching can be effectively tuned by $V_G$ and the second is that the dependence of $I_C$ on $V_G$ can be designed by the oxidation time. In particular, the second principle allows us to mimic p-type or n-type semiconductor device by defining a SOT device as p-type (n-type) when the device switches its magnetization by an in-plane input clock current $I_{IN}$ for a negative (positive) $V_G$.

**Programmable logic operation of spin logic device**

With these working principles, we first demonstrate a logic operation using a single device in which the gate voltage ($V_G$) and the input current ($I_{IN}$) are used as two input parameters (Fig. 2a). The $t_{ox}$ is 125 sec of the device (> $t_{ox.op}$; over-oxidation condition), which is thus p-type where $|I_C|$ for a negative $V_G$ is smaller than that for a positive $V_G$ (Fig. 2b). By setting the in-plane input clock current $I_{IN}$ of ±13 mA and $V_G$ of -24 V, the magnetization switches from the UP magnetization (//+z) to the DOWN magnetization (//-z) or vice versa, depending on the polarity of $I_{IN}$. Figure 2c shows the operation of the given spin logic device, which allows us to construct a truth table (Fig. 2d). Here, we define the positive values of the inputs ($V_G$ = +24 V, $I_{IN}$ = +13 mA) as digital input "1" whereas their negative values ($V_G$ = -24 V, $I_{IN}$ = -13 mA) as digital input "0". Those inputs determine the output of the spin logic device, the magnetization direction which is monitored by measuring anomalous Hall resistance $R_{xy}$; UP magnetization ($R_{xy}$ > 0) is equivalent to digital output "1" and DOWN magnetization ($R_{xy}$ < 0) is equivalent to digital output "0". Figure 2d demonstrates that two logic operations are possible depending on the initial magnetization state; the device operates the OR function of $I_{IN}$ and $V_G$ when it is initialized as "1" state (magnetization UP), and the AND function of $I_{IN}$ and $V_G$' (NOT $V_G$) when initialized as "0" state (magnetization DOWN). This offers the realization of multi-functional or programmable spin logic devices.

The logic operations can be also performed by the combination of two spin logic devices with the same type. The sample is fabricated by defining two nominally identical CoFeB/MgO/AlO$_x$ islands on a common Ta underlayer, which are designated as spin logic devices A and B, respectively (Fig. 3a). We note that both devices A and B are of p-type ($t_{ox}$ = 125 sec) and $|I_C|$ is 12.5±0.5 mA (14.5±0.5 mA) for a $V_G$ -24 V (+24 V) [Fig. 3b]. Figure 3c shows the performance of the device for initial UP magnetization of both devices A and B ($R_{xy,A} = R_{xy,B} = +2\Omega$), where the $I_{IN}$ and $V_G$ is set to ±13 mA and -24 V, respectively. The magnetization or the $R_{xy,A\ (B)}$ of the device A (B) is controlled by the $I_{IN}$ only when $V_{G,A\ (B)}$ of -24 V is applied, confirming the p-type character. Using the results for the initial UP magnetization and $I_{IN}$ of -13 mA extracted from Figs. 3b and 3c, we construct a truth table (Fig. 3d), where two input parameters of $V_{G,A}$ and $V_{G,B}$ [the positive (negative) $V_G$ corresponds to the digital input "1" ("0")] determine the output of the device, magnetization configuration which is monitored by the sum of the Hall resistances ($R_{output} = R_{xy,A}+R_{xy,B}$). In order to describe the logic operation, we define the reference, $R_{ref,\ UP\ (DOWN)} = \frac{R_{xy,A}+R_{xy,B}}{2}$, for UP (DOWN) magnetizations in both devices: $R_{ref,\ UP}$=+2$\Omega$ and $R_{ref,\ DOWN}$=-2$\Omega$. The truth table shows that the $R_{output}$ is larger than the $R_{ref,\ UP}$ corresponding to the digital output "1", only when both $V_{G,A}$ and $V_{G,B}$ are positive, demonstrating the AND gate operation. On the other hand, the OR gate can be performed by taking the $R_{ref,\ DOWN}$; the $R_{output}$ is smaller than the $R_{ref,\ DOWN}$ corresponding to the digital output "0", only when both $V_{G,A}$ and $V_{G,B}$ are negative. Note that the different logic operation can be attained by changing the initial magnetization and the polarity of $I_{IN}$ as shown in Supplementary Note 3.

**Complementary functionality of spin logic devices**

The next demonstration is for the complementary operation of the spin logic devices, in which n-type device ($t_{ox}$ = 25 sec; under-oxidation condition) and p-type device ($t_{ox}$ = 125 sec; over-oxidation condition) are electrically connected and incorporated with a common gate electrode $V_G$ (Fig 4a). The n-type characteristic of the former device is shown in Supplementary Note 4. Before the operation, we initialize the devices to be UP magnetizations ($R_{xy,n} = R_{xy,p}$ = +2Ω) corresponding to the digital output "1" state and use the $I_{IN}$ of ±12 mA. Figure 4b shows the complementary operation of the spin logics; when we apply $V_G$ = +24 V (red-shaded region), the magnetization or digital state of the n-type device is manipulated by the $I_{IN}$ while that of the p-type device remains unchanged. In contrast, when we apply $V_G$ = -24 V (blue-shaded region), the digital state in the p-type device is selectively controlled by the $I_{IN}$. Therefore, our devices perform the n-type (p-type) function exclusively for the positive (negative) $V_G$, confirming the complementary functionality of the spin logic device. This is a spin analogue to the conventional CMOS, which can facilitate the spin-based logic devices. Furthermore, when we apply an $I_{IN}$ greater than 14 mA, both devices are simultaneously controlled regardless of the gate voltage (grey-shaded region). This is a unique function of the spin logic devices based on SOT, which allows for initializing or erasing all the information at the same time [38].

**Discussion**

We finally discuss several advantages of our spin logic device over conventional CMOS-based device: (i) it does not require refreshing in quiescent mode because of its non-volatile characteristic of magnetic state as well as VCMA effect (Fig. S2), resulting in a substantial reduction in power consumption. (ii) The n-type or p-type character of the spin logic device is determined by the oxidation state at the FM/oxide interface, which in principle can be controlled by a gate voltage for an improved VCMA effect. This reconfigurable spin logic device offers great flexibility for various logic algorithms. (iii) The device does not include semiconductors, meaning that three-dimensional integration of the devices would be possible, which can result in considerably increased device density. (iv) It can also serve as a built-in memory cell within a logic circuit, reducing power dissipation required for transferring signals between commonly separate logic and memory cells. (v) The switching of the SOT-based device is fast [39], which are prerequisites for logic operations. We note that the signal of our device can be considerably improved by introducing a magnetic tunnel junction to read the output of the device [40, 41]. This together with above advantages promises that the further development of our spin logic device could provide an efficient logic unit with high density and fast speed.

## Methods

**Sample preparation** The samples of Ta(5 nm)/Co$_{32}$Fe$_{48}$B$_{20}$(CoFeB, 1 nm)/MgO(1.6 nm)/AlO$_x$(1.8 nm) structure were prepared by magnetron sputtering on thermally oxidized Si substrates with a base pressure of less than $4.0 \times 10^{-6}$ Pa ($3.0 \times 10^{-8}$ Torr) at room temperature. All metallic layers were grown by DC sputtering with a working pressure of 0.4 Pa (3 mTorr), while MgO layer is deposited by RF sputtering (150 W) from an MgO target at 1.33 Pa (10 mTorr). AlO$_x$ was formed by depositing 1.5 nm of a metallic Al layer followed by exposing to O$_2$ plasma with a pressure of 4 Pa (30 mTorr) and a power of 30 W for various oxidation times $t_{ox}$. In order to promote the perpendicular magnetic anisotropy (PMA), samples were annealed at 250 °C for 40 min in vacuum condition. The Hall-bar structured devices including a square-shaped ferromagnetic island were fabricated using photo-lithography and Ar ion-beam etching. The width of the Hall bar is 5 μm or 10 μm and the size of the ferromagnetic island varies from $4 \times 4$ μm$^2$ to $10 \times 10$ μm$^2$. The gate oxide of ZrO$_2$ (40 nm) was deposited by atomic layer deposition (ALD) at 150 °C using JUSUNG SDP ALD system.

**Measurements** The anomalous Hall resistance ($R_{xy}$) was measured using a DC current of 100 μA while sweeping in-plane magnetic field ($B_x$), parallel to the current ($I_x$) direction, under various gate voltages ($V_G$). The spin-orbit torque (SOT) switching experiments were carried out by applying a pulsed current ($I_x$) of 20 μs and a fixed in-plane magnetic field $B_x$ of 10 mT under $V_G$'s. The logic operations were performed by controlling magnetization direction using a pulsed input clock current ($I_{IN}$) of 20 μs and a gate voltage of ±24 V under an in-plane field ($B_x$) of 10 mT. All measurements were carried out at room temperature.

**Data availability** Authors can confirm that all relevant data are included in the paper and/or its supplementary information files and data are available on request.

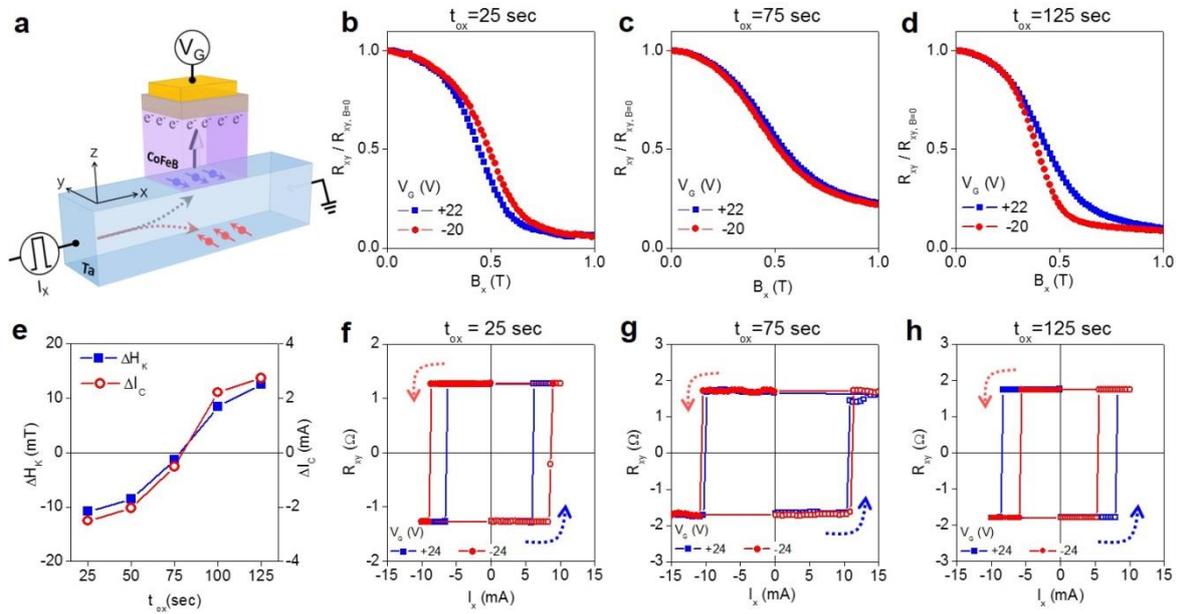

**Figure 1 | Manipulation of magnetic anisotropy ($H_K$) and critical switching current ($I_C$) by gate voltage ($V_G$). a**, Schematic of the sample with Ta(5 nm)/CoFeB(1 nm)/MgO(1.6 nm)/AlO$_x$ (1.8 nm) structures, where the magnetization direction can be controlled by the combination of in-plane current ($I_x$) and gate voltage ($V_G$). **b,c,d**, Voltage-controlled magnetic anisotropy (VCMA) for samples with oxidation time ($t_{ox}$) of 25 sec (**b**), 75 sec (**c**), and 125 sec (**d**), respectively. The normalized anomalous Hall resistance $R_{xy}/R_{xy, B=0}$ is plotted as a function of in-plane magnetic field ($B_x$) under a $V_G$ of +22 V or -20 V. Here $V_G$ of 20 V corresponds to an electric field of 5.0 MV/cm. **e**, The change in $H_K$ and $I_C$ for samples with various $t_{ox}$'s by the $V_G$, where $\Delta H_K$ and $\Delta I_C$ are $H_K$ ($V_G$= +22V) - $H_K$ ($V_G$= -20V) and $I_C$ ($V_G$= +24V) - $I_C$ ($V_G$= -24V), respectively. **f,g,h**, Spin-orbit torque (SOT)-induced switching experiments for samples with $t_{ox}$ = 25 sec (**f**), $t_{ox}$ = 75 sec (**g**), and $t_{ox}$ = 125 sec (**h**). The $R_{xy}$ vs in-plane current $I_x$ under a $V_G$ of +24 V or -24 V. During the switching experiments, a magnetic field of 10 mT is applied along the direction of the $I_x$.

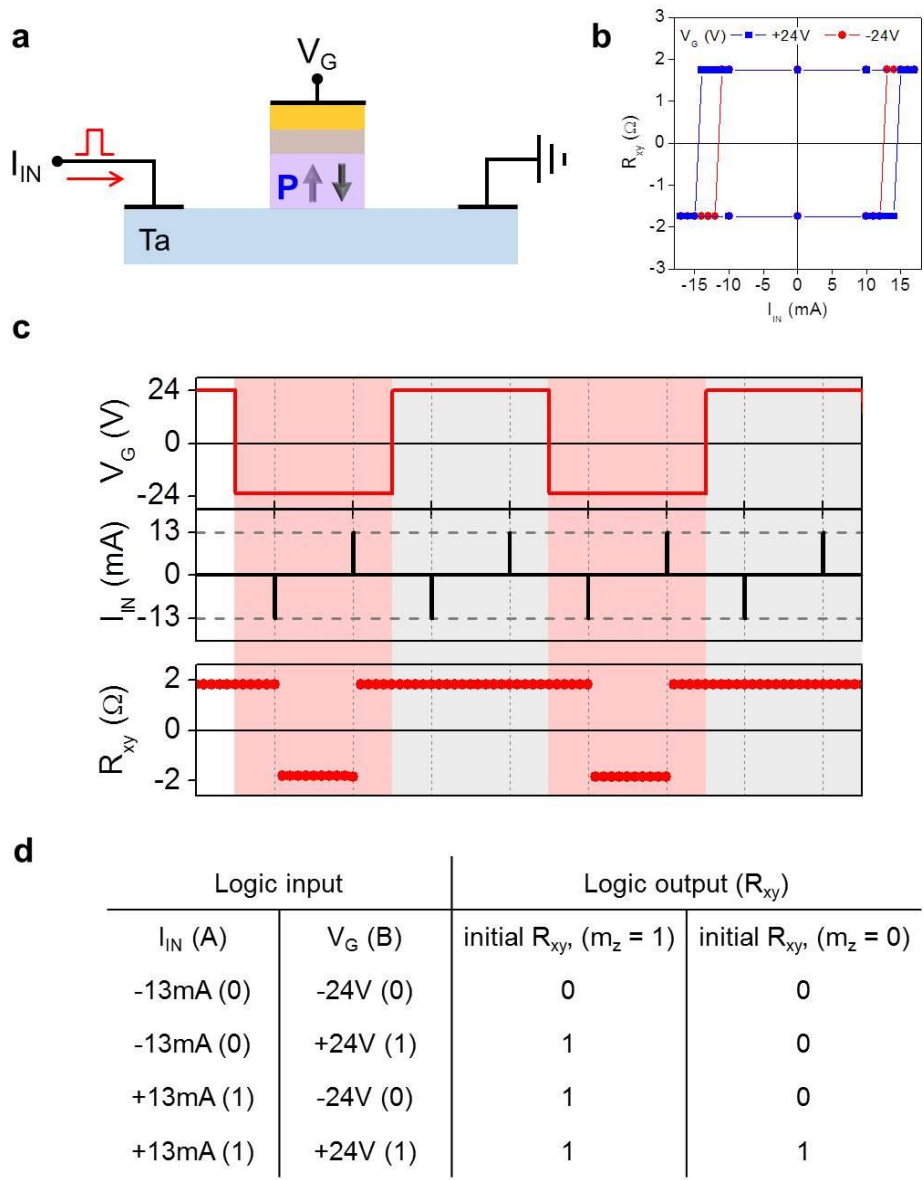

**Figure 2 | Logic operation of a single spin logic device**. **a**, Schematic of the device, in which the gate voltage ($V_G$) and the input current ($I_{IN}$) are used as two input parameters. **b**, SOT-induced switching characteristic of the spin logic device under the $V_G$'s of ±24 V. **c**, Input parameters and output of the device. Upper two panels: sequence of the $V_G$ (= ±24 V) and the $I_{IN}$ (= ±13 mA). Lower panel: The anomalous Hall resistance $R_{xy}$ representing the magnetization direction that is determined by the $V_G$ and $I_{IN}$. **d**, Logic true table constructed using the results in Figure 2c.

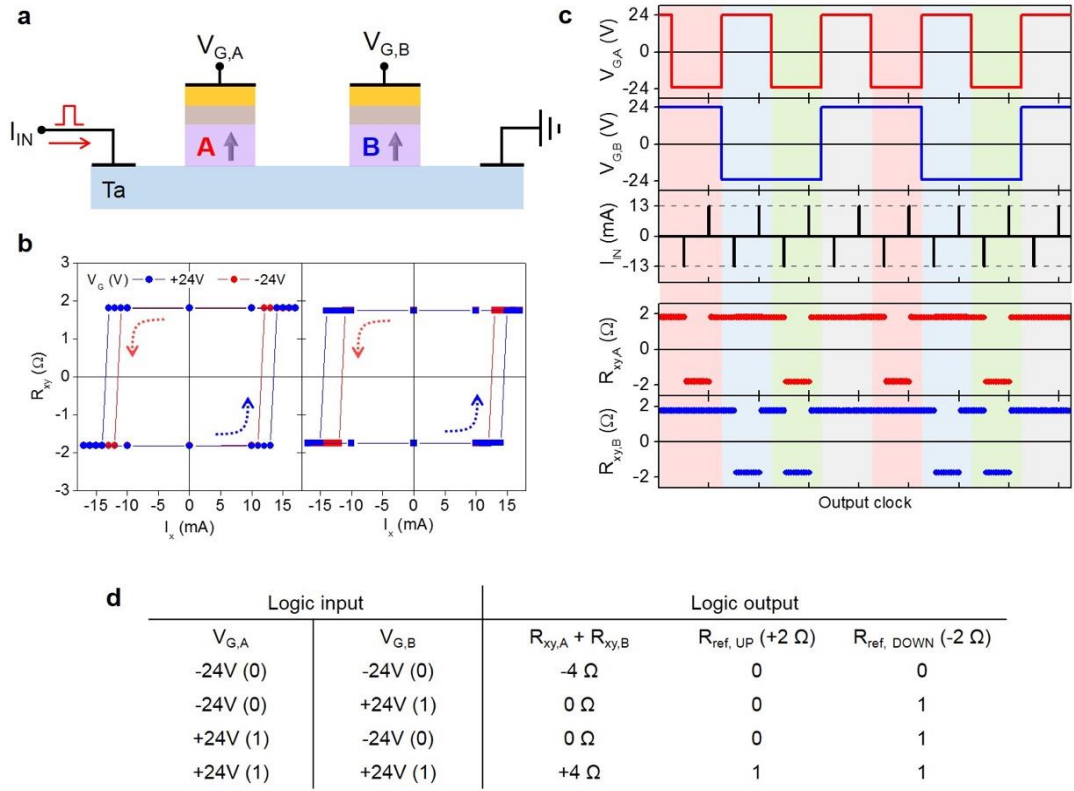

**Figure 3 | Logic operation of p-type spin logic devices**. **a**, Schematic of the device, consisting of two p-type spin logic devices which are denoted as spin logic device A and B, respectively. **b**, SOT-induced switching characteristic of device A (left panel) and B (right panel) under the $V_G$'s of ±24 V. **c**, Input parameters and output of the device. Upper three panels: sequence of the $V_G$ (= ±24 V) and the $I_{IN}$ (= ±13 mA). The $V_{G,A}$ ($V_{G,B}$) is the $V_G$ applied on device A (B). Lower two panels: The anomalous Hall resistance $R_{xy}$ representing the magnetization direction that is determined by the $V_{G,A}$ ($V_{G,B}$) and $I_{IN}$. The $R_{xy,A}$ ($R_{xy,B}$) is the $R_{xy}$ of device A (B). **d**, Truth table of the logic operations in the spin logic devices: input parameters of $V_{G,A}$ and $V_{G,B}$, output of the device $R_{output}$ (= $R_{xy,A}+R_{xy,B}$), and the reference $R_{ref,\,UP\,(DOWN)}= \frac{R_{xy,A}+R_{xy,B}}{2}$, for UP (DOWN) magnetizations of both devices. Both devices are initialized to have UP magnetization directions and $I_{IN}$ = -13 mA

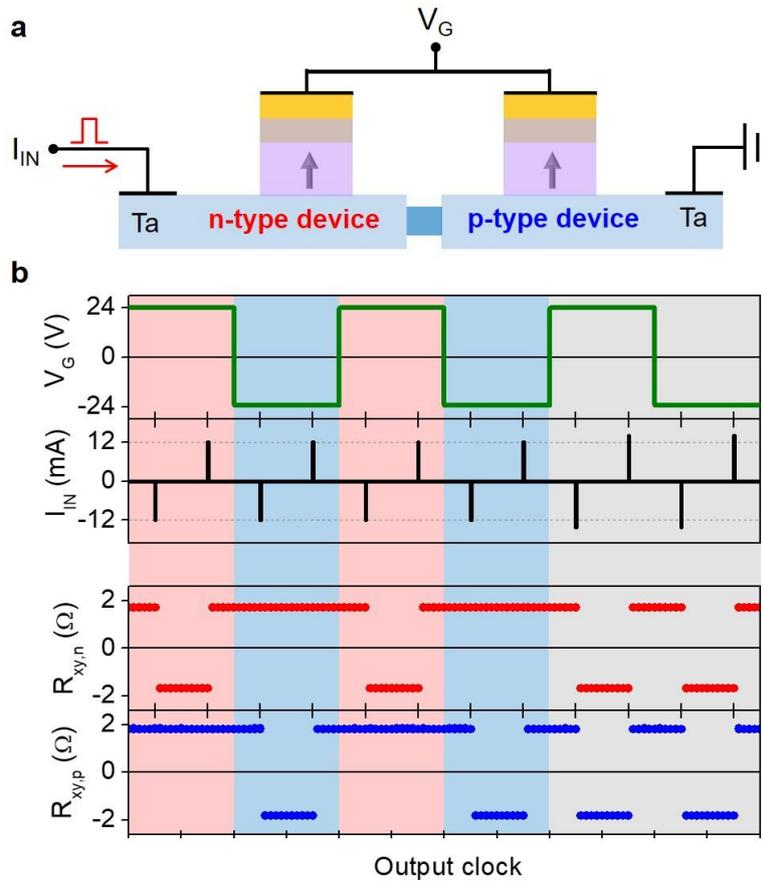

**Figure 4 | Demonstration of complementary functionality in spin logic devices**. **a**, Schematic of the device consisting of n-type and p-type spin logic devices, which are electrically connected and incorporated with a common gate electrode $V_G$. **b**, Input parameters and output of the device. Upper two panels: sequence of the input parameters: common gate voltage, $V_G$ = +24 V (red-shaded region), -24 V (blue-shaded region), and $I_{IN}$ = ±12 mA. The grey-shaded region is for $I_{IN}$ = ±14 mA. Lower two panels: The anomalous Hall resistance $R_{xy}$ representing the magnetization direction that is determined by the $V_G$ and $I_{IN}$. The $R_{xy}$ of n-type and p-type devices are denoted as $R_{xy,\,n}$ and $R_{xy,\,p}$, respectively.

**Supplementary Note 1. Dependence of magnetic anisotropy on oxidation time**

Figures S1a-S1e show the anomalous Hall resistance ($R_{xy}$), representing the z-component of magnetization ($m_z$), as a function of the in-plane magnetic field ($B_x$) for the Ta(5 nm)/CoFeB(1 nm)/MgO(1.6 nm)/AlO$_x$(1.8 nm) samples with different oxidation times $t_{ox}$'s. The x-component of the magnetization, $m_x$ ($=\sqrt{1-m_z^2}$) versus $B_x$ is shown in the inset of each graph, from which the $H_K$ is extracted (Fig. S1f). Note that all the measurements are done without a gate voltage (as-deposited state). We find that the $H_K$ is the largest for $t_{ox} = 75$ sec, referring to as the optimal oxidation state, $t_{ox,op}$, and $H_K$ decreases for $t_{ox}$ either longer or shorter than $t_{ox,op}$.

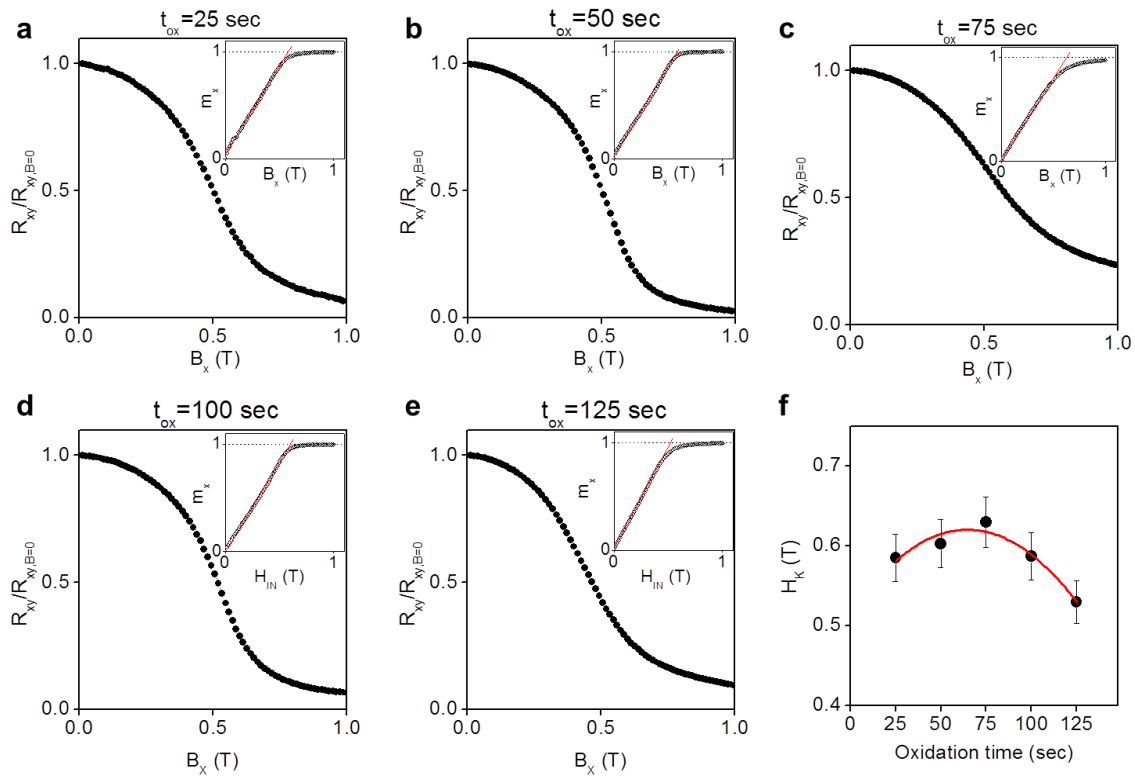

**Figure S1 | Magnetic anisotropy field ($H_K$) versus oxidation time ($t_{ox}$). a,b,c,d,e** Normalized anomalous Hall resistance ($R_{xy}$) as a function of in-plane magnetic field ($B_x$) for Ta(5 nm)/CoFeB(1 nm)/MgO(1.6 nm)/AlO$_x$(1.8 nm) samples of $t_{ox}=$ 25 sec (a), 50sec (b), 75 sec (c), 100 sec (d), and 125 sec (e), respectively. Insets show $m_x$ ($=\sqrt{1-m_z^2}$) versus $B_x$ curves. **f**, Summary of the $H_K$ versus $t_{ox}$. The red line is a guide to the eye.

**Supplementary Note 2. Non-volatility of voltage-controlled magnetic anisotropy**

As shown in Fig. 1 of the main text, the magnetic anisotropy field ($H_K$) of the over-oxidized sample ($t_{ox}$ = 125 sec) decreases by a gate voltage $V_G$ of -20 V, which is demonstrated in Fig. S2, where the black (red) symbols represent the normalized $R_{xy}$ versus $B_x$ curve under 0 V (-20V). We repeated the same measurement after removing the $V_G$, which is indicated as the blue symbols in Fig. S2, demonstrating the $H_K$ remains the same as the previously-measured value under -20 V. This clearly shows non-volatile behavior of the voltage-controlled magnetic anisotropy in our samples.

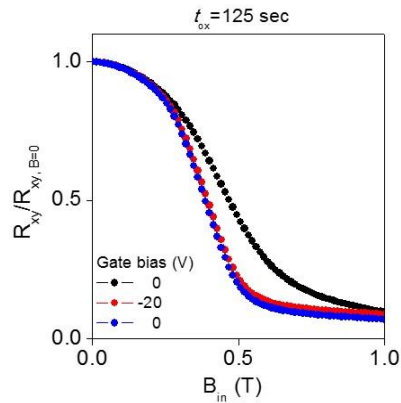

**Figure S2 | Non-volatility of voltage-controlled magnetic anisotropy.** Normalized anomalous Hall resistance versus in-plane magnetic field for the over-oxidized sample ($t_{ox}$ = 125 sec) under a sequence of gate voltage, 0 V (black symbols), -20 V (red symbols), and 0 V (blue symbols).

**Supplementary Note 3. Logic operation of spin logic device**

Table S1 summarizes the logic operation of the device of Fig. 3 on the main text when both devices are initialized as DOWN magnetization and the input clock current is positive ($I_{IN} > 0$). By changing the initial magnetization states, the same device exhibits different Boolean equations from those in Fig. 3d of the main text; demonstrating that the logic operation can be programmed by the initial magnetization direction.

**Table S1.** Logic operation of two p-type spin logic devices for initial magnetization DOWN and the positive $I_{IN}$.

| Logic input | | Logic output | | |
|---|---|---|---|---|
| $V_{G,A}$ | $V_{G,B}$ | $R_{xy,A} + R_{xy,B}$ | $R_{ref,UP}$ (+2 Ω) | $R_{ref,UP}$ (-2 Ω) |
| -24V (0) | -24V (0) | +4 Ω | 1 | 1 |
| -24V (0) | +24V (1) | 0 | 0 | 1 |
| +24V (1) | -24V (0) | 0 | 0 | 1 |
| +24V (1) | +24V (1) | -4 Ω | 0 | 0 |

**Supplementary Note 4. Electric-field controlled spin-orbit torque switching in n-type spin logic device**

Figure S3 shows the dependence of the spin-orbit torque (SOT) switching on the gate voltage $V_G$ for the n-type device ($t_{ox}$= 25) in the complementary spin logic devices presented in Fig. 4 of the main text. The $I_C$ in the n-type spin logic device increases (decreases) when a negative (positive) $V_G$ of 24 V is applied, which is the opposite to that in p-type device, which is summarized in Table S2.

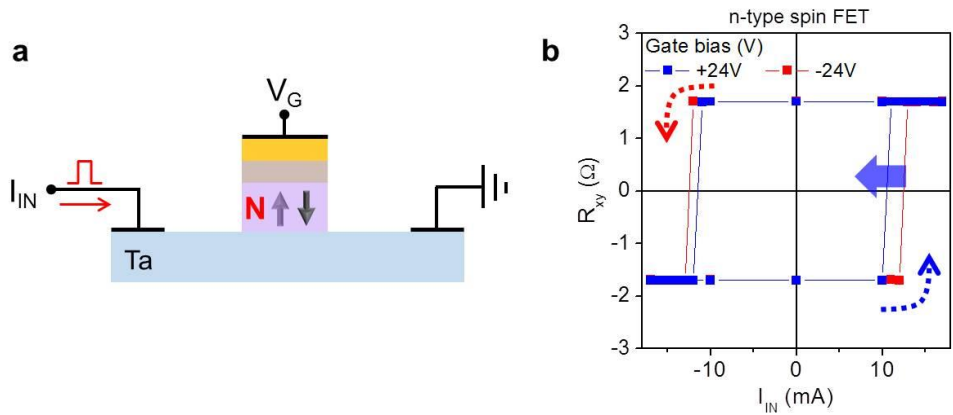

**Figure S3 | Electric-field controlled spin-orbit torque switching. a**, Schematic of n-type ($t_{ox}$= 25) devices. **b**, SOT switching characteristics of the devices under a positive and negative $V_G$ of 24 V.

**Table S2 |** Summary of critical current of SOT-induced switching for the complementary spin logic device consisting of n-type and p-type devices shown in Fig. 4 of the main text.

|  | $I_{C, \text{n-type}}$ | | $I_{C, \text{p-type}}$ | |
|---|---|---|---|---|
| $V_G$ | +24V | -24V | +24V | -24V |
| UP→DOWN | -12mA | -13mA | -14mA | -12mA |
| DOWN→UP | +11 mA | +13mA | +14 mA | +12mA |